\documentclass[preprint,aps,showpacs,preprintnumbers]{revtex4}
\usepackage{amsmath}

\usepackage{graphicx}
\usepackage{dcolumn}
\usepackage{bm}

\begin{document}

\title{Photonic mode density effects on single-molecule fluorescence blinking}
\author{F. D. Stefani}
\altaffiliation{Currently at Institute of Photonic Sciences - ICFO, Castelldefels
(Barcelona), Spain}
\email{fernando.stefani@icfo.es}
\author{K. Vasilev}
\author{N. Bocchio}
\author{F. Gaul}
\author{A. Pomozzi}
\author{M. Kreiter}
\email{kreiter@mpip-mainz.mpg.de}
\affiliation{Max Planck Institute for Polymer Research, Ackermannweg 10, D-55128 Mainz,
Germany}
\date{\today}

\begin{abstract}
We investigated the influence of the photonic mode density (PMD) on the
triplet dynamics of individual chromophores on a dielectric interface by
comparing their response in the presence and absence of a nearby gold film.
Lifetimes of the excited singlet state were evaluated in ordet to measure
directly the PMD at the molecules position. Triplet state lifetimes were
simultaneously determined by statistical analysis of the detection time of
the fluorescence photons. The observed singlet decay rates are in agreement
with the predicted PMD for molecules with different orientations. The
triplet decay rate is modified in a fashion correlated to the singlet decay
rate. These results show that PMD engineering can lead to an important
suppression of the fluorescence, introducing a novel aspect of the physical
mechanism to enhance fluorescence intensity in PMD-enhancing systems such as
plasmonic devices.
\end{abstract}

\pacs{32.50.+d, 73.20.Mf, 78.67.-n}
\maketitle


%\section{introduction}
The rate of spontaneous photon emission by an excited molecule can
be modified by changing the density of possible electromagnetic
decay channels, i.e. the photonic mode density (PMD)
\cite{Purcell1946, Barnes1998, Lakowicz2003} at the emitter
position. In particular, a nearby metallic object supporting
plasmonic excitations can produce noticeable changes in the PMD.
Together with the increased optical excitation in a locally
enhanced electric field, modifications of the PMD leading to a
faster de-excitation have been identified as responsible for the
enhanced fluorescence signal close to metal structures. This
approach to obtain stronger fluorescence signals promises
significant progress for experimental schemes whrere weak
fluorescence signals need to be retrieved with high
signal-to-noise ratios such as single molecule methods,
fluorescence correlation spectroscopy and biosensors
\cite{Dulkeith2005,Levene2003}. Following first experiments in
front of a metallic mirror
\cite{Drexhage1968,Kuhn1970,Chance1978}, recent progress in
nanofabrication technologies allowed to investigate more
sophisticated geometries showing more drastic modifications of the PMD \cite%
{Ditlbacher2001,Rigneault2005,Farahani2005}.Recently, a number of investigations addressed the quantitative analysis of fluorescence in PMD modifying environments \cite{Anger2006, Buchler2005, Enderlein2005}. In all these investigations, the PMD effects were analyzed in the framework of an enhanced singlet decay. In this paper, we show that the PMD can also influence the triplet state dynamics (fluorescence blinking) of chromophores, leading to a significant additional fluorescence enhancement mechanism.

\begin{figure}[tbp]
\includegraphics{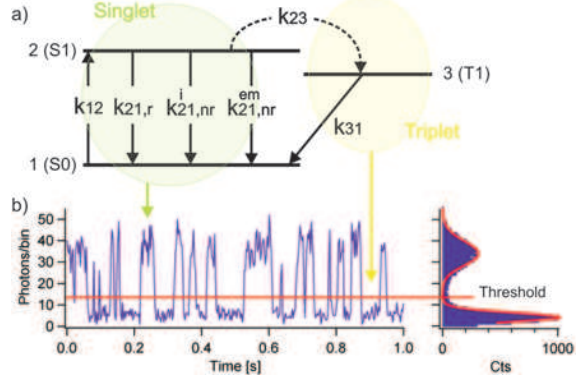}
\caption{a) Three-state description of the fluorescence process of a single
molecule and the rates $k_{i}$ involved: $k_{12}$ excitation, $k_{21,r}$
radiative de-excitation, $k_{21,nr}$ non-radiative intramolecular
de-excitation, $k_{21,nr}^{em}$ electromagnetic non-radiative de-excitation,
and the triplet population and depopulation rates $k_{23}$ and $k_{31}$. b)
1\thinspace s time trace of the fluorescence emission of a single DiI
molecule next to the experimental distribution of photons per bin (grey) and
a double-Poisson distribution (black) with averages equal to the on- and
off-intensities.}
\label{fig_3ls_rates}
\end{figure}
FIG.\thinspace \ref{fig_3ls_rates} depicts a simplified picture of the
molecular fluorescence process in terms of a three-state model \cite%
{Moerner1997}. A molecule initially in the ground singlet state S0 is
excited upon photon absorption to the singlet excited state S1 with a rate $%
k_{12}$. From S1, the molecule returns to S0 with a total rate $k_{21}$
composed of a radiative decay $k_{21,r}$ and an intramolecular non-radiative
relaxation $k_{21,nr}$; based on previous studies \cite%
{Vasilev2004,Stefani2005}, we will consider $k_{21,nr}$ as negligible. A
nearby metallic object can modify $k_{21,r}$ and introduces additional
electromagnetic non-radiative channels $k_{21,nr}^{em}$. Efficient
fluorophores perform many cycles like this in the singlet subspace while
fluorescence photons are emitted. Although suppressed by spin selection
rules, singlet$\rightarrow $triplet and triplet$\rightarrow $singlet
transitions (inter-system crossing ISC) still occur. An excited molecule has
a finite probability ($k_{23}$) of undergoing ISC to a lower energy,
long-lived triplet state (T1) from which the molecule decays back to S0 with
a rate $k_{31}$; this latter step may involve photon emission
(phosphorescence). As a result, the fluorescence emission of single
molecules is intermittent \cite{Basche1995}. If the triplet lifetime is long
enough (small $k_{31}$) this triplet blinking can be clearly observed in the
fluorescence emission of single molecules as successive bright (on) and dark
(off) periods (FIG.\thinspace \ref{fig_3ls_rates}b), each characterized by
its fluorescence and background intensity, respectively. Although ISC does
not usually affect the overall fluorescence quantum yield due to the low
absorption of T1 at the frequency of the $S0\rightarrow S1$ excitation, it
limits the maximum achievable fluorescence intensity \cite{Basche1995}.
A reduction of the triplet lifetime of organic fluorophores, with a corresponding increase in brightness, was demonstrated
by the presence of molecular oxygen which quenches T1 and returns the
molecules to S0 up to 100 times faster\cite{English2000,Huebner2001}. However, oxygen is undesired as
it is also responsible for most photodegradation processes.

Although most pronounced PMD enhancements are observed on irregular metal
structures such as silver island films or sharp metallic tips or junctions,
we used planar systems which are better suited for a quantitative study
of the underlying physical effects because they can be prepared with high
accuracy and can be fully modeled.
\begin{figure}[tbp]
\includegraphics{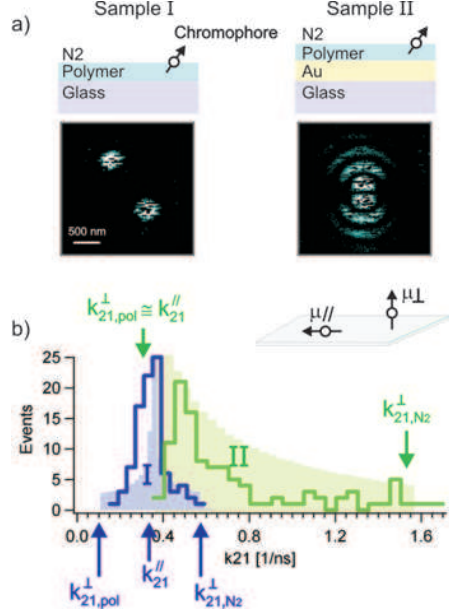}
\caption{Experimental. a) Schematic representation of the samples and
fluorescence (epi-illumination from the glass side) micrographs of single
molecules in each sample. b) Experimental distributions and calculated
values of singlet excited state decay rates for samples I and II.}
\label{fig_samples}
\end{figure}

Two sample architectures were employed. Firstly, chromophores were placed on
top of a dielectric layer (Samples I). Secondly, the same dielectric layer
was used as a spacer to place the chromophores at a controlled distance from
a thin gold film (Samples II). FIG.\,2a shows a schematic of the two sample
architectures and corresponding single molecule fluorescence confocal
micrographs (epi-illumination through the glass slides) \cite{Stefani2005};
the fluorescence blinking can be directly observed. The thicknesses of the
gold film and the dielectric spacer layer were 44\,nm and 30\,nm,
respectively in order to achieve optimum single molecule detection \cite%
{Stefani2005, Vasilev2004}. The dielectric spacer was composed of
alternating layers of poly(allylamine) (PAH) and poly(styrenesulfonate)
(PSS). On the samples with the gold film, the polyelectrolyte layers were
prepared following a published method \cite{Stefani2005}. The samples
without the gold film were prepared using the same procedure to deposit 2.5
PSS/PAH bilayers on optically transparent glass substrates, which were
previously functionalized with 3-aminopropyltriethoxy-silane (3-APTES) \cite%
{SI}. On both types of samples, fluorescent 1,1$^{\prime }$,3,3,3$^{\prime }$%
,3$^{\prime }$-hexamethylindicarbocyanine iodide (DiIC1(5),\,\emph{Molecular
Probes}) molecules were deposited electrostatically on the negatively
charged surface of the multilayer terminated with PSS by immersing the
samples in a $10^{-10}\,$M, Milli-Q water solution. The immersion time ($%
\approx$10-30\,seconds) was adjusted such as to obtain well separated
chromophores. The polyelectrolytes were chosen as dielectric spacer because
they can be prepared with controlled thickness and allow for the
chromophores at their interface to adopt all possible orientations \cite%
{Stefani2005}. Chromophores of the DiI family are a well investigated model
system for ISC effects \cite{English2000, Huebner2001} mainly due to their
relatively long triplet lifetime of 10-50 ms (in the dry state and absence
of O$_2$). All measurements were conducted under continuous flow of dry
nitrogen in order to exclude triplet quenching by oxygen.

A region of the samples was imaged on a home-built confocal microscope, in
an epi-illumination scheme from the glass side, a single molecule moved into
the focus, and its fluorescence emission recorded as a function of time.
Light from a laser-diode (\emph{Hamamatsu PLP10-063/C}) was used for pulsed
excitation (100\,ps FWHM, 100\,MHz repetition rate) at $\lambda=633\,$nm
through a 1.4\,NA oil-immersion objective. In order to optimize the signal
to background ratio, annular illumination was used for samples II \cite%
{Stefani2005}. Fluorescence photons were collected by the same objective,
separated from the excitation light by suitable dichroic and notch filters
and their arrival times were recorded by means of an avalanche photo-diode (%
\emph{Perkin-Elmer SPCM-AQR-13}) and a time correlated single photon
counting module (\emph{Becker\,\&\,Hickl SPC-630}). For each detected photon
two times were recorded independently: the time elapsed since the last
excitation pulse (micro-time) and the time elapsed since the beginning of
the measurement (macro-time), with a resolution of 7\,ps and 50\,ns,
respectively. The excited singlet state decay $k_{21}$ was obtained by a
single-exponential fit to a histogram of the micro-times \cite{SI}. The
system response was limited to approx. 0.6\,ns by the response of the
photodetectors \cite{SI}. Information about the fluorescence blinking is
provided by the macro-times, which permit to obtain the fluorescence intensity vs.time trace as shown in FIG.\,1 and in turn the length of the on- and off-times. The probabilities of a certain on- or off-time length are exponentially distributed \cite{Basche1995, Huebner2001,SI} and therefore, average times $\tau_{on}$ and $\tau_{off}$ can be extracted
which are connected to the transition rates in the 3-level system (FIG.\,\ref%
{fig_3ls_rates}) by $\tau_{on}=(k_{21}+k_{23})/(k_{12}k_{23})$ and $%
\tau_{off}=1/k_{31}$. Alternatively, $\tau_{on}$ and $\tau_{off}$ can also be obtained from an exponential fit to the fluorescence intensity autocorrelation \cite{Bernard1993}. The latter method turns out to be more reliable, especially at low signal to background conditions \cite{SI}.

Changes in the PMD affect all transitions coupled to the electromagnetic
field. This effect is well understood for $k_{21}$ and can be fully modeled
by considering point oscillating electric dipoles interacting classically
with the electromagnetic field \cite{Anger2006,Buchler2005,Barnes1998}. In the following, an analysis of the variations of $k_{21}$ is presented which gives direct access to the PMD experienced by the individual molecules. This information is a prerequisite for the analysis of the PMD-induced variations of the triplet dynamics that is shown later.

For chromophores near a plane interface, $k_{21}$ depends on the polar angle $%
\phi$ between the dipole moment of the molecule and the surface normal and
can be calculated from the limiting cases of molecules parallel ($\phi=\pi/2$%
) and perpendicular ($\phi=0$) to the interfaces:
\begin{equation}
k_{21}(\phi)= \sin^2({\phi})\ k_{21}^{\parallel}\ +\ \cos^2({\phi})\
k_{21}^{\bot}  \label{eq_emission_arb_orient}
\end{equation}
$k_{21}^{\parallel}$ and $k_{21}^{\bot}$ are determined by the dielectric
properties of the layered system \cite{Vasilev2004}.

Due to the very short immersion times used in the sample preparation, the
chromophores cannot diffuse inside the polymer \cite{vonKlitzing1996} and
are placed at the interface. In point-dipole theory, $k_{21}^{\parallel}$
varies continuously across the interface whereas $k_{21}^{\bot}$ takes two
different values at each side of the interface according to the dielectric
contrast:
\begin{align}
k_{21,N_2}^{\parallel} = k_{21,pol}^{\parallel}\ \ ;\ \
\epsilon_{N_2}\,k_{21,N_2}^{\bot} = \epsilon_{pol} k_{21,pol}^{\bot}
\label{eq_k21_disc}
\end{align}
with the subscripts $N_2$ and $pol$ referring to an infinitely small
displacement to the nitrogen and polymer side, respectively. On a molecular
level, the chromophores have a finite size and the polymer surface has a
certain structure leading to different radiative decay rates for molecules that probe more one or the other side of
the interface. Thus, a distribution of $k_{21}$ due to molecular orientation and local environment may be anticipated.

In FIG.\ref{fig_samples}b, the experimental distributions of $k_{21}$
obtained from approximately 100 molecules in each type of samples are
displayed, together with calculated distributions. Arrows indicate the theoretical decay
rates for parallel and perpendicular dipoles at each side of the interface
\cite{Vasilev2004}. These calculated rates need to be adjusted to the
experimental ones by one commom scaling factor for all calculated rates
\cite{Stefani2005}; i.e. a reference is needed. We assign the most frequent value of $k_{21}$ in samples
I to parallel molecules because those are the statistically most probable
and most effectively detected molecules. The adequacy of this choice is supported by the very good agreement
between the calculated and experimentally detected extreme values of $k_{21}$ for molecules in samples I and the maximum detected $k_{21}$ for molecules in samples II. We know from previous studies that in samples II, parallel molecules and molecules probing more the polymer side of the interface are not detectable in this scheme \cite{Stefani2005}. Then, the minimum observed $k_{21}$ is expected to be larger than the minimum calculated $k_{21}$ ($k_{21}^{parallel}$).
The calculated distributions of $k_{21}$ shown in FIG.\,\ref{fig_samples}b correspond to randomly oriented molecules, half of them placed in N2 and half in the polymer. Comparing the calculated and experimental distributions of $k_{21}$, it is possible to conclude that the sample preparation yields a smaller fraction of molecules with extreme lifetimes than expected from random orientations and location; i.e. more molecules lie parallel to the surface and/or probe the polymer side of the interface. In summary, the variations of the PMD has been quantified by means of $k_{21}$ for both samples and its distribution can be fully explained by the different location and orientation of the molecules with respect to the interface and, in samples II, to the nearby gold film.

\begin{figure}[tbp]
\includegraphics{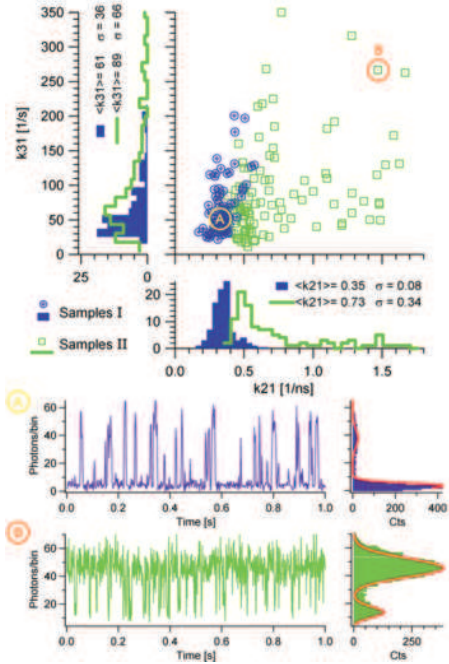}
\caption{Distributions of $k_{31}$ and $k_{21}$ and scatter plot of $k_{31}$%
\,vs.\,$k_{21}$ for 91 molecules in samples\,I and 108 molecules in
samples\,II. Below, 1-s fluorescence emission traces of an average molecule
in samples\,I (A) and a strongly enhanced molecule in samples\,II (B). Next
to the traces, the photons-per-bin histograms are shown.}
\label{fig_k31VSk21}
\end{figure}

With this information in hand, we proceed to investigate the effect of the PMD on $k_{31}$. In FIG.\,%
\ref{fig_k31VSk21}, the distributions of $k_{21}$ and $k_{31}$ obtained from
molecules in samples I and II are shown together with a scatter plot of $%
k_{31}$ vs $k_{21}$. The distributions of $k_{31}$ in samples I and II
differ significantly (Student-t test: t=3.25, P=0.14\%): Molecules in
samples II have both an increased average $k_{31}$ and large $k_{31}$ values
exceeding $200\,\mathrm{s^{-1}}$ that are not observed in the absence of the
gold film. In spite of some statistical scatter, a positive correlation
between $k_{31}$ and $k_{21}$ can be observed in both data sets. Pearson
linear correlation coefficients of $R=0.46$ and $R=0.35$ are obtained for
the molecules in samples I and II, respectively. The significance of these
correlations was tested by computing the probability of obtaining by chance
such $R$ values or higher from non-correlated $k_{31}$ and $k_{21}$ data.
Probabilities of $0.006\%$ and $0.035\%$ for samples I and II, respectively
were obtained \cite{SI}. The correlation between $k_{21}$ and $k_{31}$ can
be explained by PMD variations for the individual molecules due to their
local environment and orientation if $k_{31}$ depends in a similar fashion
on the PMD as does $k_{21}$. While some interaction with the surroundings on
a molecular scale affecting both transitions could also produce such a
positive correlation, this can be excluded as the reason for any difference
between the molecules in samples I and II because they experience
the same chemical environment in both samples. Therefore, the significant
difference between the distributions of $k_{31}$ in samples I and II must be
assigned to a PMD-mediated enhancement due to the nearby gold film. This in
turn supports the interpretation of the positive correlation between $k_{21}$
and $k_{31}$ in terms of similar PMD effects on the singlet and triplet
de-excitations.

The noticeable effect of the PMD on $k_{31}$ indicates that the transition T1%
$\rightarrow$S0 occurs mainly radiatively and that the transition dipoles
associated with the T1$\rightarrow$S0 and S1$\rightarrow$S0 transitions have
similar orientations. The latter is reasonable because for ISC to occur S1
and T1 should share the same molecular geometry. It is important to note
that the PMD enhancements achieved in this planar geometry are quite modest
in comparison to what can be achieved in more complex structures. The
enhanced $k_{31}$ can lead to noticeable stronger fluorescence signals, due
to the faster return to the singlet manifold which allows to obtain more
fluorescence photons per time unit. As illustrative examples, two 1-second
traces of an average (A) and a strongly enhanced (B) molecule are shown also
in FIG.\,\ref{fig_k31VSk21} (An intermediate case is shown in FIG.\,\ref%
{fig_3ls_rates}). Clearly, molecule B is a much brighter emitter than A because it spends much more time in the singlet
subspace emitting fluorescence photons.

%\section{summary}
In conclusion, the influence of the PMD on the electronic transition rates
involved in molecular fluorescence was investigated by studying the emission
blinking and excited state lifetime of individual molecules. The singlet
de-excitation rate is affected in good agreement with theory. It was
demonstrated that the triplet de-excitation is affected in a similar way by
the PMD indicating that for the investigated system the T1$\rightarrow$S0
transition has a strong radiative component with a transition dipole moment
of similar orientation to the S1$\rightarrow$S0. These findings complete the
picture of the PMD-mediated fluorescence enhancement which was so far only
discussed for the singlet manifold. Firstly, an enhanced excitation ($k_{12}$%
) and singlet decay ($k_{21}$) at constant branching rate to the triplet ($%
k_{23}$) allows the dye to emit more photons before entering the triplet and
secondly, the residence time in the triplet is reduced due to its enhanced
decay ($k_{31}$). This new finding on the improvement of chromophore
performance should encourage further the investigation of PMD enhancing
structures such as plasmonic nano-objects to obtain single super emitters.

\begin{acknowledgments}
This work was supported by the Bundesministerium f\"ur Bildung und Forschung
(grant No. 03N8702).
\end{acknowledgments}

\end{document}